\documentstyle[aps,12pt]{revtex}
\begin{document}
\tightenlines
\title{Classical and quantum time dependent solutions in string theory}
\author{C. Mart\'{\i}nez-Prieto, O. Obreg\'on and J. Socorro }
\address{Instituto de F\'{\i}sica de la Universidad de Guanajuato,\\
Apartado Postal E-143, C.P. 37150, Le\'on, Guanajuato, M\'exico\\
}
\maketitle
\begin{abstract}
Using the ontological interpretation of  quantum mechanics in a particular 
sense, we obtain the classical behaviour of the scale factor and two scalar 
fields, derived from a string effective action for the FRW time dependent 
model. Besides, the  
Wheeler-DeWitt equation is solved exactly.  We speculate that the same 
procedure could also be applied to S-branes.
\end{abstract}


\section{Introduction}

There are several attempts to understand diverse aspects of time 
dependent models in the framework of string theory and branes 
[for a review see Ref. \cite{quevedo}].
In this context, new interesting results arise:  Starting with an 
$AdS_5$ metric one can specify the localization of a brane and by 
these means a 4D FRW cosmology emerges on the brane \cite{langver},  
D-Brane inflation has also been proposed \cite{dvali},  the 
interaction between D-branes and anti-D-branes lead to the study of Tachyon 
condensation and its role in inflation \cite{sen}.  The intersection of  
branes at different angles \cite{arashei} generates an inflation potential.
The rolling tachyon has also been considered in connection with inflation 
\cite{alexmaz}.  Compactified M-theory on an interval $S^1/Z_2$ and further compactified 
in a six-dimensional Calabi-Yau leads to two 4-D worlds at the ends
of the interval in the 5D bulk.  It turns out that besides boundary 
branes, those at the end of the intervals, there are also 5-D  branes
 that can move through the bulk.  A very interesting proposal was made 
 \cite{khoury} assuming that a bulk brane going from one boundary of the interval to the 
  other end, would collide with the second boundary brane and produce the 
  big bang.  It has been claimed that the density fluctuations measured 
  in the CMB could be a consequence of quantum fluctuations of the bulk 
  brane. Also, there is no need for an inflation potential, an attractive 
potential 
  is proposed  describing the attraction of the branes and as a consequence 
  the scale factor depends on the position of the brane in the interval.  
  This model has, however, been very much debated in the literature \cite{kallkhou}.

On the other hand, it is well known that relativistic theories of 
gravity such as general relativity or string theory are invariant under
reparametrization of time. The quantization of such theories presents a number
of problems of principle loosely known as ``the problem of time'' 
\cite{isham,kuchar,zeh}. This problem occurs in all system whose classical
version is invariant under reparametrization of the time parameter, which
leads to the absence of this parameter at the quantum level. The formal 
question is how to handle the classical Hamiltonian constraint, $\rm H 
\approx 0$, in the quantum theory. Also, connected with the problem
of time is the ``Hilbert space problem'' \cite{isham,kuchar}, it is not
at all obvious which inner product of states one has to use in quantum
gravity, and whether there is a need for such a structure at all. 

In some situations, the notion of time can be recovered in quantum cosmology.
Basically, the approaches to address the problem of time can be classified as
to whether an appropriate time variable is identificable already at the 
classical level, or only after quantization. The former try, for example,
to cast the hamiltonian constraint through an appropriate canonical
transformation into the reduced form $\rm P_T + H \approx 0$, where
$\rm P_T$ denotes the momentum conjugate to the time variable T or, in other
words $\rm P_T$ is the momentum associated with  a variable, that appears 
linearly in the hamiltonian constraint, and after quantization this
implies a Schr\"odinger equation.

If this were possible, the problem of time would be solved,
since the new form of the constraints would be transformed into a
Schr\"odinger type of equation upon quantization, and a standard inner product
could be proposed to tackle the Hilbert space problem. However, it is known that
this cannot work for the full configuration
space \cite{torre}.  One uses then a Klein-Gordon type of inner product \cite{witt}, since this
inner product is not positive definite,  many problems arise
which have lead some authors to invoke a ``third quantization'' of the
theory.

The purpose of this paper is to present a mechanism to obtain  classical solutions 
for time dependent models 
from the quantum regime, without having to solve directly
the field equations of motion. The idea is that when the behaviour of the 
wave function with respect to the gravitational part leads to an 
{\it increasing} function, this type of solution cannot correspond to a classical 
trajectory. Therefore, there is a forbidden region in the evolution of our 
universe. Then, we need a wave function having a decreasing behaviour with
respect to the scale factor A.  The behaviour 
of the scale factor may be determined in 
two different forms from the quantum regime:  we can
calculate the expectation value of the scale factor, in the spirit
of the many worlds interpretation of quantum mechanics \cite{world}
\begin{equation}\rm
<A>_t= \frac{\int_0^\infty {\cal W}(A) \, \Psi^*(A,t) \,A(t)\, \Psi(A,t)\, dA}
{\int_0^\infty {\cal W}(A) \, \Psi^*(A,t) \, \Psi(A,t) \, dA},
\end{equation}
where $\rm {\cal W}(A)$ is a weight function that normalizes the 
expectation value. The other way corresponds to use  the  WKB semiclassical
approximation, evaluating the Bohmian trajectories in the
ontological formulation of quantum mechanics \cite{bohm},  where the system 
follows a real trajectory given by the equation
\begin{equation}
\rm \Pi_q= \frac{\partial \Phi}{\partial q},
\label{wkb}
\end{equation}
where the index q designates one of the degrees of freedom of the
system, and $\Phi$ is the phase of the wave function 
\begin{equation}
\rm \Psi = W \, e^{i\Phi},
\end{equation}
where W and $\Phi$ are real functions. 

A classical solution was found using the last approach for a gravity model  
coupled to barotropic perfect fluid, as matter field, and cosmological 
constant\cite{socorro}. We will show that
the solution found for the cosmological model under consideration, in a particular effective
 string theory, has the same kind of behaviour than in ref. \cite{socorro}.  

On the other hand, recently a new class of objects were introduced in 
 string theory named spacelike or S-brane \cite{gutperle}, objects that
exist only for a moment of time. The main motivation for the introduction of 
the S-branes was the conjectured dS/CFT correspondence \cite{strominger}. 
The expectative is that, using
the analogy with p branes, S-branes can also be found as explicit
solutions of Einstein equations (coupled to scalar fields),  the
S-branes solutions are then {\it time-dependent} backgrounds of the theory.
Therefore, the approach we shall follow for a cosmological model could also 
be analyzed for a S-brane solution  (see \cite{leblobuch} and references therein).

We will consider the canonical quantization of a low-energy string effective 
action for a particular cosmological metric.  This procedure is interesting 
by its own right and being the quantization (for a model dependent on time) 
of the effective theory of strings one would expect to gain some information
 on the quantization of the objects depending on time in string theory.

We begin by considering the compactification of the NS-NS sector of string
effective action which contains the dilaton field, the graviton and a 2-form
potential and is common to both type II and heterotic theories. Consider the
form of the (D+d)dimensional NS-NS action compactified on T$^{d}$ (d torus).
In (D+d) dimensions, the action is given by

\begin{eqnarray}
S&=&\int d^{D+d}x\sqrt{ g_{D+d}}
e^{-\Phi }\left[ R_{D+d}(g)+ g^{AB} \nabla_{A}\Phi 
\nabla_{B} \Phi  -\right. \nonumber\\
&& \left.  -\frac{1}{12} H_{M_1 M_2 M_3} H_{M_{1^{\prime}}
M_{2^{\prime}} M_{3^{\prime }}} g^{M_1 M_{1^{\prime }}} g^{M_2 M_{2^{\prime}}}
g^{M_3 M_{3^{\prime }}}\right] .
\end{eqnarray}
In our case we compactify on 6-torus $T^6$ with internal metric
\begin{equation}
\rm h_{ab}= diag \left( e^{-2\sigma}, e^{-2\sigma}, e^{-2\sigma}, 1, 1, 1
\right).
\label{hab}
\end{equation}
where $a,b= (4,5,6,7,8,9)$. Choosing $H_{456}=F=constant.,$ and by compactifing 
the coupling parameter $\rm \Phi$ becomes
\begin{equation}
\rm \Phi = 2 \phi - \frac{1}{2} ln (det \, h_{ab}),
\end{equation}
where $\rm d=6$, $\rm D=4$. In the internal space we have
$\rm h_{44}=h_{55}=h_{66}=e^{-2\sigma}$, thus
by means of straightforward calculations and choosing for simplicity
 $\rm F=0$,
we arrive to an action in the
string frame  with two scalar fields (dilatonic and moduli field). 

\begin{equation}\rm 
S=\frac{1}{2\kappa^2}\int d^4 x \sqrt{-g} e^{-2\phi -3\sigma} 
\left[R+4 g^{\mu\nu}
\partial_\mu \phi \partial_\nu \phi 
+ 12 g^{\mu\nu}\partial_\mu \phi \partial_\nu \sigma + 6 g^{\mu\nu} 
\partial_\mu \sigma \partial_\nu \sigma  \right].
\label{action1}
\end{equation}
The static model with  
$\rm H^{\mu \nu b}\not=0$ has been considered in the literature \cite{kuri}
in relation  with black hole solutions. In \cite{copeland}
 exact solutions for homogeneous, anisotropic cosmologies in
four dimensions were obtained for the low-energy string effective action
including a homogeneous dilaton $\phi$ and antisymmetric tensor field 
$\rm B_{\mu \nu}$ (axion field).

Under  the dynamical compactification scheme, one expect that the radius 
depending on the
scalar field $\sigma$ should be  a small constant or vanishing, in such
a way that the time dependence of $\sigma$, keeps $\sigma$ large.

The standard Friedmann-Robertson-Walker (FRW) metric is
 
\begin{equation}\rm 
 ds^2=- N^2 (t) dt^2 + A^2 (t) \left[\frac{dr^2}{1-\kappa r^2}
+r^2\left(d\theta^2 + sin^2\theta d\phi^2\right)\right],
\label{metric}
\end{equation}
where $\rm N$ is the lapse function, $\rm A$ is the scale factor of the model,
and $\kappa$ is the curvature index of the universe
( $\rm \kappa=0,+1,-1$ flat, close and open, respectively). From this 
metric, the scalar curvature becomes
\begin{equation} \rm
 R=-\frac{6\ddot A}{AN^2} + \frac{6\dot N\dot A}{N^3A}- \frac{6k}{A^2}
- 6 \left(\frac{\dot A}{NA}\right)^2 ,
\end{equation}
where the overdot denotes time derivative. 
Thus, the action (\ref{action1}) is written as

\begin{eqnarray}
\rm S&=&\rm\frac{1}{2\kappa^2} \int d^3 x \sqrt{ g}
\int dt \quad e^{-2\phi -3\sigma}  
\left\{ 6 \left[ - \left(\frac{A^2 \dot A}{N}  \right) ^{^\bullet} + 
\frac{A(\dot A)^2}{N} -NkA \right]  \right. \nonumber\\ 
& -& \rm \left. \frac{A^3}{N} \left[ 4 \dot\phi^2
+  12 \dot\sigma \dot\phi 
+ 6 \dot \sigma^2 \right] \right \} .
\label{action2}
\end{eqnarray}

In the Einstein frame, we perform the conformal transformation into the
metric components 

 \begin{equation}\rm 
\bar g_{\mu\nu} = e^{-(2\phi + 3\sigma )} g_{\mu\nu}, 
\end{equation}
getting
\begin{eqnarray}
\rm N &=& \rm e^{\phi +\frac{3}{2}\sigma} \bar N ;\qquad 
A= e^{\phi +\frac{3}{2}\sigma}
\bar A \nonumber \\
\rm \dot A &=& \rm e^{\phi+\frac{3}{2}\sigma} \left[ \dot{\bar{A}} 
+ (\dot\phi +
\frac{3}{2} \dot\sigma ) \bar A \right],
\label{conformal}
\end{eqnarray}
and substituting the equations (\ref{conformal}) in the action
(\ref{action2}), we get

\begin{equation}\rm 
S=\frac{1}{2\kappa^2} \int d^3 \, x \sqrt{\bar g} \int dt  \left\{ \left(
\frac{6\bar A\dot{\bar{A}}^2}{\bar N}-6k \bar N \bar A\right)
- \frac{\bar A}{\bar N}^3 \left(10 \dot\phi^2 + 30\dot\phi\dot\sigma +
\frac{39}{2} \dot\sigma^2 \right) \right\}.
\label{action3}
\end{equation}

Given (\ref{action3}), the work is organized as follows. In the next section 
we will get the 
Hamiltonian, using the procedure given by
Arnowitt-Deser-Misner (ADM),  obtaining the Wheeler-DeWitt (WDW)
equation for the three coordinate fields, in Sec. III we solve this 
WDW equation. The classical time behaviour of the coordinate fields is discussed in 
Sec. IV,  the solutions are written in quadratures with respect to
the scale factor, due to the fact  that the solution was obtained 
in terms of elliptic integrals
of first and second class. We take  the particular case of a flat
 universe, where the solutions are found in a closed way. Sec. V is devoted 
 to remarks.

\section{The Hamiltonian}
The Lagrangian density  of our model is 
\begin{equation}
{\cal L} = \rm \frac{6A\dot A}{N}^2 -6k NA-\frac{A^3}{N} \left( 10 \dot\phi^2
+30\dot\phi \dot\sigma +\frac{39}{2} \dot\sigma^2 \right) ,
\label{lagrangian1}
\end{equation}
where the bars have been omitted. 
The canonical momenta to coordinate fields are defined  in the usual way
\begin{eqnarray}
\rm  \Pi_A &\equiv & \frac{\partial{\cal L}}{\partial\dot A} =
\rm \frac{12A\dot A}{N},\nonumber\\
\rm \Pi_\phi &\equiv & \frac{\partial{\cal L}}{\partial\dot\phi}= 
\rm-\frac{A^3}{N} (20\dot\phi + 30\dot\sigma ), \label{momentosigma}\\
\rm \Pi_\sigma &\equiv  &\frac{\partial{\cal L}}{\partial\dot \sigma} 
=-\frac{A^3}{N} (30\dot\phi + 39\dot\sigma ),\nonumber
\end{eqnarray}
from which we can obtain the temporal derivative of $\rm \phi$ and $\rm \sigma$
fields

\begin{eqnarray}
\rm \dot\sigma &=& \rm \frac{N}{A^3} \left(- \frac{3}{12}\Pi_\phi 
+ \frac{\Pi_\sigma}{6} \right),\label{ssigma}\\
\rm \dot\phi &=& \rm \frac{N}{A^3} \left(\frac{13}{40} \Pi_\phi 
-\frac{1}{4} \Pi_\sigma \right) .
\label{pphi}
\end{eqnarray}

Using (\ref{momentosigma}, \ref{ssigma}, \ref{pphi}), we can rewrite the 
Lagrangian density (\ref{lagrangian1}) in the canonical form

\begin{equation}
{\cal L}=\rm \Pi_A \dot A+\Pi_\phi \dot\phi +\Pi_\sigma \dot\sigma - N \left\{
\frac{\Pi^2_A}{24A}+6kA +\frac{13}{80}\frac{\Pi^2_\phi}{A^3} +
 \frac{\Pi^2_\sigma}{12A^3}-\frac{1}{4}\frac{\Pi_\sigma\Pi_\phi}{A^3}
\right\}.
\label{lagra} 
\end{equation}

We can see that the Hamiltonian  is given by

\begin{equation}
\rm  H=    \frac{\Pi^2_A}{24A} +6kA
+\frac{13}{80}\frac{\Pi^2_\phi}{A^3}
+\frac{\Pi^2_\sigma}{12A^3}-\frac{1}{4} \frac{\Pi_\phi \Pi_\sigma}{A^3}
  .
\label{hami}
\end{equation}

Performing the variation of N in the Lagrangian density (\ref{lagra}), we
get the equation $\rm H=0$.  Following the standard procedure to get the 
quantum version of this
Hamiltonian, we promote to operators the canonical momenta $\rm \Pi_\lambda$
that satisfy the commutation relation $\rm [\Pi_\lambda, \lambda]=-i$
with the representation 
$\rm \Pi_\lambda= -i \frac{\partial}{\partial \lambda}$. We then apply 
the operator $\rm \hat H$ to the wave function $\rm \Psi$, i.e. 
$\rm \hat H \Psi =0$.

Notice that in principle the  ambiguity order in equation 
(\ref{hami}) should be taken into account. This is quite a 
difficult problem to be treated in all its generality. We will,
however, consider the proposal in \cite{HH} (there are other possibilities depending on 
different considerations on the operators, see refs. \cite{za1,li1})

\begin{equation}
\rm \Pi^2_A = -A^{-B} \frac{\partial}{\partial A}  A^{B} 
\frac{\partial}{\partial A} 
=-\frac{\partial^2}{\partial A^2}  
-\frac{B}{A}\frac{\partial}{\partial A}, 
\end{equation}
where B is a real parameter that measure this ambiguity. For the other 
operators this problem does not arise
\begin{equation}\rm
\Pi^2_\phi =- \frac{\partial^2}{\partial\phi^2} \,\, ; \,\,\qquad\qquad
\Pi^2_\sigma =- \frac{\partial^2}{\partial\sigma^2} .
\end{equation}
 In this way (\ref{hami}) can be written as
\begin{equation}\rm
 \frac{1}{24A} \left(-\frac{\partial^2}{\partial A^2} -\frac{B}{A} 
\frac{\partial}{\partial A} \right) \Psi + 6kA\Psi -\frac{13}{80}
\frac{1}{A^3} \frac{\partial^2}{\partial\phi^2} \Psi - \frac{1}{12A^3}
\frac{\partial^2}{\partial\sigma^2}\Psi 
 +\frac{1}{4A^3} \frac{\partial^2}{\partial \phi \partial \sigma}\Psi =0, 
\label{hami1}
\end{equation}
and multiplying  (\ref{hami1}) by $\rm 24A^3$, we obtain the equation

\begin{equation}\rm
- A^2\frac{\partial^2}{\partial A^2}\Psi -B A \frac{\partial}{\partial A}
\Psi +144 k A^4 \Psi -\frac{39}{10} \frac{\partial^2}{\partial\phi^2} \Psi 
- 2\frac{\partial^2}{\partial\sigma^2}\Psi 
+6\frac{\partial^2}{\partial\phi\partial\sigma} \Psi =0.  
\label{13}
\end{equation}

\section{Quantum solution}
Employing the separation of variables method to solve (\ref{13}),
 $\rm \Psi (A,\phi ,\sigma ) = {\cal A}(A) {\cal C} (\phi ,\sigma)$
and rearranging, we get
\begin{equation}\rm
 \frac{1}{{\cal A}}  \left(-A^2 \frac{d^2 {\cal A}}{dA^2}-B A 
\frac{d {\cal A}}{d A} +144 k A^4 {\cal A} \right) + 
 \frac{1}{{\cal C}} \left(-\frac{39}{10} 
\frac{\partial^2 {\cal C}}{\partial\phi^2} 
-2\frac{\partial^2 {\cal C}}{\partial\sigma^2}
+6\frac{\partial^2 {\cal C}}{\partial\phi\partial\sigma} \right) =0 . 
\end{equation} 
This equation is equivalent to the set of partial differential equations
\begin{eqnarray}
\rm A^2 \frac{d^2{\cal A}}{dA^2} +B A \frac{d{\cal A}}{dA}- 
\left(144 kA^4 \pm \nu^2\right){\cal A} &=&\rm 0 ,\label{bessel} \\
\rm \frac{39}{10} \frac{\partial^2 {\cal C}}{\partial \phi^2}
+2\frac{\partial^2 {\cal C}}{\partial \sigma^2} 
-6\frac{\partial^2 {\cal C}}{\partial\phi\partial\sigma} &=&\rm 
\pm \nu^2 {\cal C},
\label{campos}
\end{eqnarray}
where $\nu$ is a separation constant.

Equation (\ref{bessel}) can be transformed to a Bessel differential 
equation for the function  $\rm \Phi$, performing the  transformations 
$\rm z=6\sqrt{-k}A^2$, ${\cal A}=\rm A^{\frac{1-B}{2}}\Phi(z)$, 
whose solution for $\rm k \not = 0$  becomes
\begin{equation}
 {\cal A}= \rm A^{\frac{1-B}{2}} Z_\alpha \left(6 \sqrt{- k} A^2 \right), 
\label{bess}
\end{equation}
where $\rm Z_\alpha$ is a generic Bessel function, with order
$\rm \alpha= \frac{1}{4} \sqrt{(1-B )^2-4(\pm \nu^2)}$. For $\rm k=1$, we
have the modified Bessel functions $\rm I_{\alpha}$ and $\rm K_\alpha$. 
For $\rm k=-1$, if $\rm \alpha$ is not integer, the solutions become
the ordinary Bessel function $\rm J_{\pm \alpha}$, in other case
we have a combination of the Bessel functions $\rm J_\alpha$ and 
$\rm Y_\alpha$ \cite{Gra} .

At this point, we need some conditions on the wave function, in such a way
 that the 
classical solutions are not forbidden for the scale factor A. Then, we need 
 a wave function having a decreasing behaviour with
respect to the scale factor A, it can be shown that this is the case
 when the parameter $B \ge 1$
and the order of the generic Bessel function $\alpha>0$; besides we choose 
that the generic Bessel function will be ($K_\alpha$ or $J_\alpha$), the 
modified
Bessel function or ordinary Bessel function, according to case. 
With these conditions 
on the parameters, we postulate that the classical behaviour will be obtained,
at least in the semiclassical approximation.

For the particular case of  $\rm k=0$, the solution for the scale factor
 has the behaviour
\begin{equation}
{\cal A}= \rm C_1 A^{\frac{1-B +\mu}{2}}+ C_2 A^{\frac{1-B -\mu}{2}},
\end{equation}
where  $\rm \mu$ is given by
\begin{equation}\rm
\mu = \sqrt{(1-B)^2-4 (\pm\nu^2 )}\not =0,
\end{equation}
with $\rm B>2+\nu$ in order to get a decreasing behaviour,
and when $\rm \mu =0$ the quantum  solution becomes
 ${\cal A}=\rm A^{\frac{1-B}{2}} (C_1 +C_2 Log \, A)$.

To solve the equation (\ref{campos}) for the fields  $\rm \phi ,\sigma$, 
we propose the following ansatz
\begin{equation}
{\cal C}= \rm G e^{m_1\phi} e^{m_2 \sigma},
\label{ansatz}
\end{equation}
where $\rm G$  is a constant, and  $\rm m_1, m_2$ are two complex parameters. 
Introducing (\ref{ansatz}) into (\ref{campos}), one gets 

\begin{equation}
\rm \frac{39}{10} m^2_1 + 2m^2_2 -6 m_1 m_2 = \pm \nu^2 , 
\label{cuadra}
\end{equation}
Solving  $\rm m_2$ as a function of $\rm m_1$, we get

\begin{equation}\rm
m_2=\frac{3}{2}m_1 \pm \frac{1}{2} \sqrt{ \frac{6}{5}m_1^2 \pm 2 \nu^2}.
\end{equation}
So, (\ref{ansatz})  can be written as

\begin{equation}
{\cal C}= \rm  e^{m_1(\phi + \frac{3}{2}\sigma)}\left(
A_1  e^{ \frac{1}{2} \sqrt{ \frac{6}{5}m_1^2 \pm 2 \nu^2}\sigma}
+ A_2  e^{- \frac{1}{2} \sqrt{ \frac{6}{5}m_1^2 \pm 2 \nu^2}\sigma} \right).
\label{c}
\end{equation}

A possible way to construct a "Gaussian state" is
\begin{eqnarray}
{\cal C}&=& \rm\int_{-\infty}^{+\infty}\left[ A_1(m_1)\sinh \left(\frac{1}{2} 
\sqrt{ \frac{6}{5}m_1^2 \pm 2 \nu^2} \, \sigma\right)
+ A_2(m_1) \cosh \left(\frac{1}{2} \sqrt{ \frac{6}{5}m_1^2 \pm 2 \nu^2}\, 
\sigma
\right) \right] \times \nonumber\\
&& \rm \times e^{m_1(\phi + \frac{3}{2}\sigma)} \, dm_1 .
\end{eqnarray}
This will be solution of  (\ref{campos}) under the ansatz (\ref{ansatz}).

In this way, a more general solution to  (\ref{13}) than (\ref{c}) is obtained 
\begin{eqnarray}
\rm \Psi (A,\phi ,\sigma ) &=&\rm  \left[ A^{\frac{1-B}{2}} Z_\alpha 
\left(6 \sqrt{- k} A^2 \right)\right]  
 \int_{-\infty}^{+\infty}\left[ A_1(m_1)
\sinh \left(\frac{1}{2} \sqrt{ \frac{6}{5}m_1^2 \pm 2 \nu^2} \, \sigma\right)
\right. \nonumber\\
&&\rm  \left. + A_2(m_1) \cosh \left(\frac{1}{2} 
\sqrt{ \frac{6}{5}m_1^2 \pm 2 \nu^2}\, \sigma
\right) \right] e^{m_1(\phi + \frac{3}{2}\sigma)} \, dm_1.
\end{eqnarray}

\section{Classical solutions a la WKB}
In the WKB approximation, we propose a solution as  
$\rm \Psi =\exp (iS(A,\phi,\sigma))$, where the 
function $\rm S=S_{1}(A)+S_{2}(\phi )+S_{3}(\sigma)$ is known as the 
superpotential function, being a real function, and fulfilling the usual following 
conditions 
\begin{equation}\rm
\left( \frac{\partial S}{\partial A}\right)^2 \gg \left| 
\frac{\partial^2S}{\partial A^2}\right| ,
\qquad \quad
\left( \frac{\partial S}{\partial \phi }\right)^2 \gg \left| 
\frac{\partial^2 S}{\partial \phi^2}\right|,  
\qquad \quad
\left( \frac{\partial S}{\partial \sigma }\right)^2 \gg 
\left| \frac{\partial^2 S}{\partial \sigma^2}\right|,
\label{condiciones}
\end{equation}
in this way, the Einstein-Hamilton-Jacobi equation (EHJ) is obtained

\begin{equation}
\rm A^2\left( \frac{dS_1}{dA}\right)^2+144kA^4+\frac{39}{10}
\left(\frac{dS_2}{d\phi }\right)^2+2\left( \frac{dS_3}{d\sigma }\right)^2
-6\frac{dS_2}{d\phi }\frac{dS_3}{d\sigma }=0 .  
\label{modificada}
\end{equation}
This same equation is recovered directly, when we introduce the transformation 
on the canonical momentas in  equation (\ref{hami}) multiplied by 
$\rm 24A^3$
\begin{equation}\rm
\Pi_A=\frac{\partial S}{\partial A}=\frac{dS_1}{dA} , \qquad\quad
\Pi_{\phi }=\frac{\partial S}{\partial \phi }=\frac{dS_2}{d\phi } ,\quad\qquad
\Pi_{\sigma }=\frac{\partial S}{\partial \sigma }=\frac{dS_3}{d\sigma }.
\label{cambio}
\end{equation}

Employing the separation of variables method, we have
\begin{equation}
\rm A^2\left( \frac{dS_1}{dA}\right)^2+144kA^4=-\frac{39}{10}\left( 
\frac{dS_2}{d\phi }\right)^2-2\left( \frac{dS_3}{d\sigma }\right)
^2+6\frac{dS_2}{d\phi }\frac{dS_3}{d\sigma }=\alpha,
\label{separacion}  
\end{equation}
where $\rm \alpha$ is a separation  constant
\begin{eqnarray}
{\rm A^2 \left(\frac{dS_1}{dA}\right)^2+144kA^4}&=&\alpha  ,
\label{factor}  \\
{\rm -\frac{39}{10}\left( \frac{dS_2}{d\phi }\right)^2
-2\left( \frac{dS_3}{d\sigma }\right)^2+6\frac{dS_2}{d\phi }
\frac{dS_3}{d\sigma} }&=&\alpha .  
\label{campos1}
\end{eqnarray}
Taking into account that the canonical momentum $\rm \Pi_A$ was defined as
 $\rm \Pi_A =\frac{12}{N}A\frac{dA}{dt}$, the equations
(\ref{cambio}) and (\ref{factor}) give us the following relation
\begin{equation}
\rm \frac{dA}{Ndt}=\frac{\sqrt{\alpha-144k A^4}}{12A^2}  .  
\end{equation}
Defining $\rm d\tau =Ndt$ as a physical time, we can rewrite the solution
for the scale factor  $\rm A$ with respect to the coordinate $\rm \tau $ 

\begin{equation}
\rm \tau =\int d\tau =\int \frac{12A^2dA}{\sqrt{\alpha-144kA^4}} . 
\label{escala}
\end{equation}
The relation between A and $\tau$, (\ref{escala}) can be expressed in 
terms of elliptic integrals \cite{Gra}, in this way by rewriting 
the integral, we have

\begin{eqnarray}
\rm \tau  &=&\rm \frac{1}{\sqrt{k}}\int_{0}^{A}
\frac{z^2 \, dz}{\sqrt{\sqrt{\frac{\alpha}{144k}}-z^2}
\sqrt{\sqrt{\frac{\alpha}{144k}}+z^2}}  \nonumber \\
&=&\rm \frac{1}{\sqrt{k}}\left( \sqrt{2}\sqrt[4]{\frac{\alpha}{144k}}\, 
E(\gamma ,r)
-\frac{\sqrt{\frac{\alpha}{144k}}}{\sqrt{2}\sqrt[4]{\frac{\alpha }{144k}}}
\, F(\gamma ,r)
-A\sqrt{\frac{\sqrt{\frac{\alpha}{144k}}-A^2}{\sqrt{\frac{\alpha }{144k}}
+A^{2}}}\right),  
\end{eqnarray}
where $\rm F$ and  $\rm E$ are the elliptic integrals of first and second 
class, respectively, and the parameters
\begin{eqnarray}
\rm \gamma  &=&\rm \arccos \left( \frac{A}{\sqrt[4]{\frac{\alpha}{144k}}}
\right) 
\frac{\sqrt{2}\sqrt[4]{\frac{\alpha}{144k}}}{\sqrt{\sqrt{\frac{\alpha}{144k}}
+A^2}} , \\
\rm r &=&\rm \frac{1}{\sqrt{2}} .
\end{eqnarray}

For the fields $\rm \phi $ and $\rm \sigma$, we  use the equation
 (\ref{campos1}), where we recognize a quadratic equation in the
associated momenta. This equation can diagonolized,  defining the parameters
$\rm x=\frac{dS_2}{d\phi}$, and 
$\rm y=\frac{dS_3}{d\sigma}$, where $\rm x$ and $\rm y$ will be taken as constants. 
The  solution  show us that
 $\rm  x$ is proportional to $\rm y$. Solving the quadratic equation 
(\ref{campos1}) we get 

\begin{equation}
\rm y=\frac{3x\pm \sqrt{\frac{6}{5}x^2-2\alpha}}{2}  .  
\end{equation}

Now, we have two integration constants ($\rm x, \alpha$), which could be 
determined with appropriate initial boundary conditions.

Using  $\rm \frac{d\phi}{dt}$ and $\rm \frac{d\sigma}{dt}$ as function of 
 $\rm x=\Pi_{\phi}$, $\rm y=\Pi_{\sigma}$ (see equations 
(\ref{ssigma}), (\ref{pphi}) and (\ref{cambio}))  and integrating, we obtain :
\begin{eqnarray}
\rm \sigma - \sigma_0&=& \rm \int d\sigma =(-\frac{3}{12}x+\frac{y}{6})
\int \frac{d\tau }{A^3(\tau)} , \label{sigma00}\\
\rm \phi- \phi_0&=& \rm \int d\phi =  (\frac{13}{40}x-\frac{y}{4})\int 
\frac{d\tau}{A^3(\tau)} . \label{phi00}  
\end{eqnarray}

For
$\rm k=0$,  a flat universe,  equation (\ref{escala}) 
is very simple, its solution is
\begin{equation}\rm 
\tau =\frac{4A^3}{\sqrt{\alpha}} .  
\label{tau}
\end{equation}
So,  the scale factor $\rm A$ has the following behaviour

\begin{equation}
\rm A=\left( \frac{\sqrt{\alpha}}{4}\right)^{\frac{1}{3}}
\tau^{\frac{1}{3}} .
\end{equation}
Introducing this result into  (\ref{sigma00}, \ref{phi00}) we have
\begin{eqnarray}
\rm \sigma -\sigma _0&=& \rm (-\frac{3}{12}x+\frac{y}{6})
\left( \frac{4}{\sqrt{\alpha}}
\right)^{\frac{1}{3}}\int \tau^{-1} \, d\tau =(-\frac{3}{12}x+\frac{y}{6})
\left( \frac{4}{\sqrt{\alpha}}\right) \ln \tau  , \label{sigma10 }\\
\rm \phi -\phi_0 &=& (\frac{13}{40}x-\frac{y}{4})
\left( \frac{4}{\sqrt{\alpha}}\right) \ln \tau  . \label{phi10}
\end{eqnarray}
In this manner, we can obtain the behaviour of the fields $\rm \phi $ and
 $\rm \sigma $ as functions of the scale factor $\rm A$, as follows

\begin{equation}
\rm \sigma -\sigma_0=\left( -\frac{3}{12}x+\frac{y}{6}\right) 
\left( \frac{4}{\sqrt{\alpha}}\right) 
\left( 3\ln A+\ln \frac{4}{\sqrt{\alpha}}\right), 
\end{equation}
\begin{equation}
\rm \phi -\phi_0=\left( \frac{13}{40}x-\frac{y}{4}\right) 
\left( \frac{4}{\sqrt{\alpha}}\right) 
\left( 3\ln A+\ln \frac{4}{\sqrt{\alpha}}\right).
\end{equation}
In a similar way, for the  cases, $\rm k=\pm 1$, we have 
\begin{enumerate} 
\item{} $\rm k=-1.$
\begin{eqnarray}
\rm \sigma -\sigma_0&=&\rm -\left( -\frac{3}{12}x+\frac{y}{6}\right) 
\frac{6}{\sqrt{\alpha}}\ln 
\left( \frac{A^2}{\frac{\sqrt{\alpha}}{12}+\sqrt{\frac{\alpha}{144}+A^4}}
\right), \label{11}\\
\rm \phi -\phi_0&=& \rm -\left( \frac{13}{40}x-\frac{y}{4}\right) 
\frac{6}{\sqrt{\alpha}} \ln \left( \frac{A^2}{\frac{\sqrt{\alpha}}{12}
+\sqrt{\frac{\alpha}{144}+A^4}} \right) . \label{12} 
\end{eqnarray}
\item{} $\rm k=1.$
\begin{eqnarray}
\rm \sigma -\sigma_0&=& \rm \left( -\frac{3}{12}x+\frac{y}{6}\right) 
\frac{6}{\sqrt{\alpha}}\ln \left( \frac{A^2}{\frac{\sqrt{\alpha}}{12}
+\sqrt{\frac{\alpha}{144}-A^4}} \right), \label{21} \\
\rm \phi -\phi_0&=& \rm \left( \frac{13}{40}x-\frac{y}{4}\right) 
\frac{6}{\sqrt{\alpha}} \ln \left( \frac{A^2}{\frac{\sqrt{\alpha}}{12}
+\sqrt{\frac{\alpha}{144}-A^4}} \right) . \label{22}
\end{eqnarray}
\end{enumerate}

\section{Final Remarks}

It is well known that the Wheeler-DeWitt cosmological equation is not an 
evolution equation and therefore the associated
quantum states do not evolve in time. A possible way to connect some 
parameters of the `quantum' WDW 
solutions with classical Einstein ones is by phenomenological restrictions 
imposed on the superpotential functions or {\it final conditions} over the 
wave 
function as we did. By these means we get a decreasing function
in the gravitational part of the wave function of the Universe. Using this
method, we find the classical behaviour for the scale factor, scalar fields
$\phi$ and $\sigma$ in terms of A. So even not knowing a time WDW quantum equation, our
physical assumptions allow us to connect the quantum behaviour with the
classical one. Under dynamical compactification condition, the moduli
 scalar $\sigma$ will be an increasing function, in such away that the radius
for the extra dimension vanish, see Eq. (\ref{hab}). On the other hand, our 
procedure applied to a time dependent object in the context of a low-energy 
string effective action (7) provides a quantization interesting in its own 
right and that is expected to provide information on the quantum objects 
depending on time in string theory. Their classical behaviour was 
 also presented.  It is a matter of future work to follow a similar 
 procedure for S-branes models.

\bigskip
\noindent {\bf Acknowledgments}\\
This work was partially supported by CONACyT grant 37851,
 PROMEP and Gto. University projects.

\end{document}